# Do 'altmetrics' correlate with citations? Extensive comparison of altmetric indicators with citations from a multidisciplinary perspective

Rodrigo Costas; Zohreh Zahedi, Paul Wouters

{rcostas, z.zahedi.2, p.f.wouters}@cwts.leidenuniv.nl

Centre for Science and Technology Studies (CWTS). Faculty of Social and Behavioral Sciences. Leiden University. PO Box 905, 2300 AX, Leiden (The Netherlands)

**Abstract**

An extensive analysis of the presence of different altmetric indicators provided by Altmetric.com across scientific fields is presented, particularly focusing on their relationship with citations. Our results confirm that the presence and density of social media altmetric counts are still very low and not very frequent among scientific publications, with 15%-24% of the publications presenting some altmetric activity and concentrating in the most recent publications, although their presence is increasing over time. Publications from the social sciences, humanities and the medical and life sciences show the highest presence of altmetrics, indicating their potential value and interest for these fields. The analysis of the relationships between altmetrics and citations confirms previous claims of positive correlations but relatively weak, thus supporting the idea that altmetrics do not reflect the same concept of impact as citations. Also, altmetric counts do not always present a better filtering of highly cited publications than journal citation scores. Altmetrics scores (particularly mentions in blogs) are able to identify highly cited publications with higher levels of precision than journal citation scores (JCS), but they have a lower level of recall. The value of altmetrics as a complementary tool of citation analysis is highlighted, although more research is suggested to disentangle the potential meaning and value of altmetric indicators for research evaluation.

## 1. Introduction

Since their introduction, social media have attracted the attention of many scholars who have integrated these sites into their daily scholarly practices. As a result of the introduction of these new tools, new possibilities of measuring the impact of scientific publications in social media have emerged (Wouters & Costas, 2012). The social web metrics or also called 'altmetrics' were first proposed in 2010 (Priem et. al. 2010a), referring to mentions of scientific outputs in social web tools such as Facebook, Twitter, blogs, news media or online reference management tools. Altmetrics aim to go further in the analysis of scientific activities, e.g. by analyzing the impact of outputs in different formats (e.g. blogs, datasets, etc.) as opposed to the analysis of only journal papers that has been the most traditional way of assessing impact of scientific outputs.

As an alternative way of measuring impact, altmetrics are aimed at complementing and improving the limitations of both traditional (i.e. bibliometrics) and web based (e.g. download and usage data) impact metrics and giving new insights to the analysis of impact (Galligan & Dyas-Correia, 2013). Although there is no exact definition for altmetrics, the concept is sometimes used as a generalization of article level metrics (http://en.wikipedia.org/wiki/Altmetrics) or 'alternative' metrics. From a more conceptual point of view, altmetrics is regarded as a subfield of informetrics and webometrics (Bar-Ilan et.al.,2012) and it has been proposed that the term 'influmetrics' (suggested by Cronin & Weaver already in 1995) is a better name instead of altmetrics (Rousseau & Ye, 2013). Altmetrics might contribute to expand the concept of scientific impact to other types of impact (e.g. societal, educational, cultural, etc. ) which are ignored by most traditional ways of impact assessment (Priem et. al.,2010a; Sud &Thelwall, 2013; Piwowar & Priem, 2013). At the same time it is expected that altmetrics can provide a better filter for finding relevant and significant publications at the level of article as the publications are assessed by a different audience (scholars, general public, etc.), this being in line with the "collaborative filtering system" proposed by Priem, Piwowar & Hemminger (2012b).

The development of the concept of altmetrics has been accompanied by a growth in the diversity of web based tools aimed to capture and track a wide range of researcher's outputs by aggregating altmetrics data across a wide variety of sources. The prevalence use of social web by scholars have also led to some studies conducted on the analysis of altmetrics and its relation or association with previous established impact metrics such as citation analysis. Most of these studies, have found correlations (low, medium and high) among altmetrics and citation scores suggesting that these two approaches are somehow related but that altmetrics might capture other types of impact than citations (see e.g. Priem, Piwowar, & Hemminger, 2012; Bar-Ilan, 2012; Zahedi, Costas & Wouters, 2012; Schlögl et. al., 2013; Thelwall et. al., 2013; Haustein et. al., 2013a; Sud & Thelwall, 2013; Haustein et. al., 2013b). Most of these studies have pointed out that it is necessary to develop more large-scale studies and to combine quantitative and qualitative approaches.

To contribute to this, in this paper we perform a large study of 718,315 publications covered in the Web of Science and from different disciplines for which we have attached altmetric indicators provided by Altmetric.com. Altmetric.com (http://www.altmetric.com/) is a commercial London-based tool that tracks, analyses and collects the online activity around scholarly outputs from a selection of online sources such as blogs, Twitter, Facebook, Google+, mainstream news outlets, media and other sources (Adie & Roe, 2013). Altmetric.com compiles all the social media attention gathered by a scientific publication in the so-called 'altmetric donut' or altmetric score. The altmetric score reflects both the quantity (the higher attention, the higher score) and quality (weighting according to different sources) of attention received by each item applying some kind of normalization[1] (both by all articles of similar age and in the same journal). Altmetric.com also provides the context for each social media mention, combined with the demographic data for Twitter mentions. Altmetric.com holds data on some 2.6 million unique papers published from July 2011 onwards and offers an open API thus allowing the possibility of collecting a wealth of impact metrics data.

For this study we have decided to use the data from Altmetric.com for several reasons:

- Robustness and stability of the data: Altmetric.com stores the data collected for every publication and keeps them over time, thus avoiding the problem of the 'volatility' of altmetric by providing a stable framework of data collection and indicators.
- Possibility of obtaining summarized altmetric indicators for individual publications: Altmetric.com provides summaries of altmetric indicators and performs some cleaning and

---

[1] The altmetric score reflects both the quantity (the higher attention, the higher score) and quality (weighting according to different sources) of attention received by each item applying some kind of control to avoid gaming.

- standardization of the data (e.g. by counting only the number of Tweets provided by unique Twitter users), although they also can provide more raw and detailed data if necessary.
- Presence of unique identifiers of publications: altmetric data are collected and summarized for publications for which unique identifiers such as DOIs, PubMed ids, arXiv ids or other handles of the publications are available. This makes the linkage of their data with other data systems easy and transparent, although not necessary free of limitations (e.g. not all publications have DOIs or PubMed ids).

In spite of the previous advantages of Altmetric.com, it is also important to mention that at the moment, any altmetric study is bound by the data providers of altmetric information (in our case Altmetric.com). As it has been previously discussed (Haustein et. al., 2013b; Wouters & Costas, 2012) data quality problems are an important issue in this field, and for this reason caution and modesty need to be regarded when discussing the results. In our case, we fully rely on the capacity of Altmetric.com on collecting robust and reliable altmetrics data. Therefore, we should consider this study only as an exploratory approach aimed to find general patterns in the presence of altmetrics across scientific publications.

The main approach of this paper is to study the presence and relationship of different altmetric measures across scientific publications and fields and their relationship with citations at a broad scale. The key question is whether there is any relationship between altmetric and citation indicators and, if that is the case, how this relationship can be characterized. In this line, we also want to test the general claim about the altmetrics of being better filters (at least better than journal indicators) in order to find relevant scientific publications. In this study, a combination of bibliometric approaches, correlation analyses and precision-recall methods have been applied in order to explore these questions. To the best of our knowledge, this is among the first comprehensive studies of altmetrics (another example is Haustein et al, 2013b) but in this case taking a multidisciplinary and multi-metric approach at such a broad scale.

*Objectives and research questions*

The main objective of this paper is to study the presence of altmetrics (excluding Mendeley[2]) among scientific papers in a particular "universe" of publications (i.e. in our case, publication from all fields covered in the Web of Science database) and to determine their relationship with citation indicators (namely the number of citations of a publication and the average impact of the publication journals). Specific attention will be paid to the ability of altmetric indicators to identify highly cited publications within the whole universe of scientific papers, in contrast to journal impact indicators. Thus, in this paper we take a similar approach as in a previous study of the recommendations of F1000 (Waltman & Costas 2013). The main research questions we seek to answer in this paper are the followings:

　　　　1. How are altmetrics scores distributed across publication years?

---

[2] The main reason to exclude Mendeley is that Altmetric.com only collects readership data when other altmetrics indicators have been detected (e.g. Twitter, mentions in blogs, etc.), as a result the counting of readership wouldn't be complete for all our sample of publications. Moreover, in the data provided by Altmetric.com for this study, readerships metrics were not included.

2. What is the general presence of Altmetrics in the Web of Science (WoS)-covered publications and across main fields of science?

3. To what extent do altmetrics correlate with the citations of publications and the impact of their journals?

4. Can altmetric scores identify highly cited publications better than journal citation scores?

## 2. Data and methodology

On 14[th] of October 2013, we downloaded from Altmetric.com a set of 1,589,440 records relating to publications with a DOI or some other publication identifiers (e.g. PubMed Id). From this dataset we found a total of 1,380,143 unique DOIs[3] that we considered as suitable for our analysis.

We matched this list of DOIs with the CWTS in-house version of the Web of Science (WoS) by DOI. As a result a total of 718,315 DOIs (52% of all the original DOIs provided by Altmetric.com)[4] were matched in our database[5]. In total only 7% of all the papers in the WoS (without any time restriction and with a DOI) got some altmetric score as covered by Altmetric.com. Based on this matching of publications, it was possible for us to perform a first general analysis on the presence of altmetric data for publications from different publication years (Figure 1).

---

[3] We cleaned some duplicate DOIs and also some DOI strings that were wrong. In the case of duplicates and following a contact with Altmetric.com, we summed all the altmetrics scores for the duplicate publications.
4 Altmetric.com targets to collect data on scholarly articles, without indicating any limitation, therefore we can assume that they collect data on any type of scientific article, therefore also publications not covered in the WoS (e.g. local journals, publication in non-English languages, etc.) thus explaining the relative high rate of unmatched publications with the CWTS database.
5 As an orientation, 30% of all the publications in 2011 in our WoS database did not present a DOI. These publications were not possible to match with the altmetric scores and therefore were excluded from the analysis. Thus, we focused on the 70% of publications from 2011 with DOI codes.

**Figure 1. Evolution of the number of WoS publications in Altmetric.com that at 14/10/2013 had received any altmetric score (blue bars: n. of publications with altmetrics; red line: share of publications with DOIs that have altmetrics; green line: share of publications in total that have any altmetrics).**

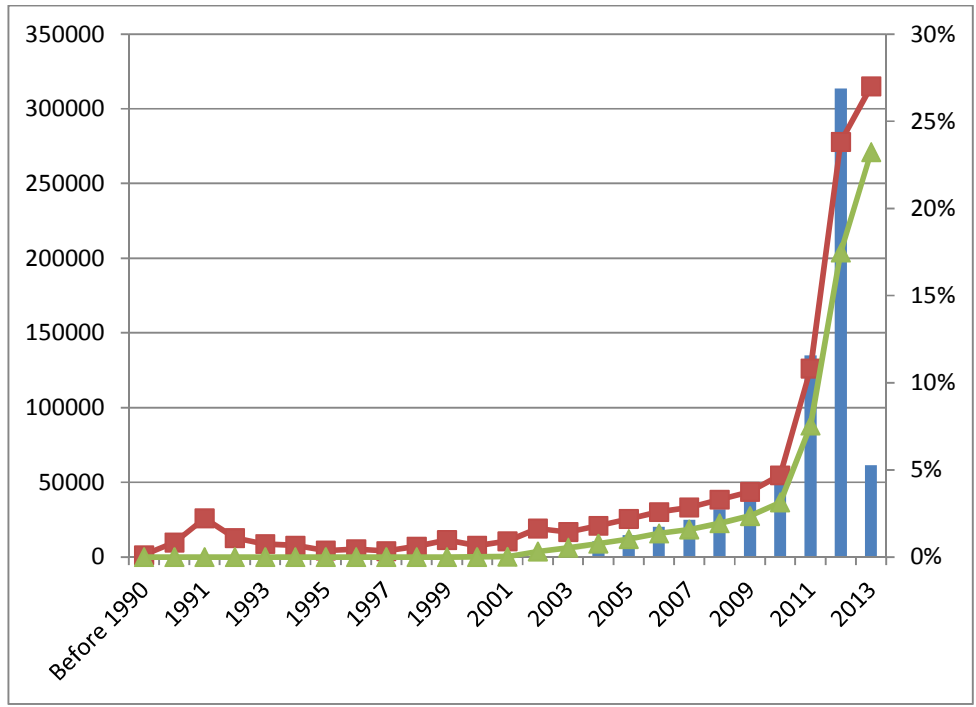

As figure 1 shows, clearly altmetric data is mostly frequent among the most recent publications, particularly 2011, 2012 and 2013. In 2011 around 10.8% of all the publications with a DOI (7.5% if we focus on all publications) received some altmetric score. This share increases to 23.8% of publications with a DOI from 2012 and above 25% in 2013. Considering all this it makes sense to highlight that altmetrics are only valid for the most recent publication years and have no real interest when applied to older publications as their presence is negligible. This is quite in line with previous studies (Haustein et. al., 2013b) that also suggest this strong 'recent bias' in altmetrics scores. Besides, it is important to take into account that Altmetric.com has started to collect data from July 2011 onwards[6], therefore publications form this moment onwards are better presented in the altmetric scores provided by this data provider.

*Filtering of publications for the study*

Given this recency bias in the altmetric indicators, we decided to focus on those publications published from July 2011 onwards. 2011 is the most recent year for which we can still have a full year of citation window (i.e. 2012) and we focused on the last months of that year given the previously mentioned limitation of Altmetric.com of a better coverage from that month onwards. However, the selection of publications from July onwards does not come without some limitations. The main limitation is that not all publications have a clear indication of their month of publication. For this reason in this study we only attributed a publication month to the paper when the month is clearly identifiable in the publication. Thus, we can say that all the publications finally selected, do actually belong to their respective months (July-December), but other publications from these months are excluded for not having clear indications of their publication months.

---

[6] Altmetric.com claims in its website that "if the article was published before July 2011, we'll have missed any transient mentions of it, tweets in particular" (http://www.altmetric.com/whatwedo.php)

A more important limitation of this focus on the last months of 2011 is that although publications get officially published in a given month and issue of a journal, it is not uncommon that they are published online in advance (e.g. through the "Online first" system even with DOIs (Wu & Notarmarco, 2003)). This time lag between the online first publication and the final official publication can be quite large (Moed, 2007; Tort, Targino & Amaral, 2012; Heneberg, 2013), thus representing a problem both for altmetrics and citations. From the altmetric point of view, this is a limitation because some of the publications in the earlier months of our population (i.e. July or August) were probably published "Online first" some months before and it could be that some of their altmetric scores were actually given before this limit of July 2011. In Appendix I, we present the evolution of publications over the months July-December 2011. As it is possible to see, although there is indeed an increase in the share of publications with altmetrics from October onwards, there is still a substantial share of publications with altmetrics scores in the two first months of our sample (having shares of publications with altmetrics around 10-12%), so we expect that the relative high number of publications included will reduce the effect of this limitation. In any case, we have introduced some complementary analytical approaches in our paper where the effect of this situation is reduced (e.g. through our so-called 'tight analysis'). Finally, in order to analyze the citations of the publications in the most robust way, we have focused only on 'article' and 'review' document types.

*Bibliometric and altmetric indicators*

For all the publications finally selected, we calculated a series of bibliometric and altmetrics indicators. The bibliometric indicators were calculated following the approach introduced by Waltman et. al. (2011) and considering a citation window of one year (2012). The following bibliometric indicators are calculated:

- ***Cs***: total number of citations received by every publication (self-citations excluded). The sum of this indicator results in the ***tcs*** (total citation score) and its average in the ***mcs*** (mean citation score) of any group of publications.
- ***Ncs***: field normalized number of citations received by every publication (self-citations excluded). For the field normalization we have considered the 250 WoS subject categories.
- ***Jcs***: journal citation score of the publication journal of the paper. The JCS of a publication in journal X equals the average number of citations received by all publications in journal X between 2011-2012. Also we calculated a second indicator named Jcs_0510_12, this is the JCS of all the publications in the same journal but published in the period 2005-2010 and counting citations up to 2012. Thus we counted with an indicator of the impact of the

- *Jfis*: this is the field normalized variant of the Jcs indicator. Also known as the Journal to Field Impact score, it is the measure of the impact of all the publications in the journal in the same period but this time normalized by the average impact of the publications in the same field (based on WoS subject categories). We have also calculated the same indicator for the period 2005-2010 with citations up to 2012 (JFIS_0510_12).
- **Top 1%** of the most cited publications (global and by major fields of Science). For some of the analysis we identified the top 1% most cited publications overall and by main fields. This identification was based on the sorting of publications based on the ncs[7] indicator, and selecting the top 1% most cited[8]. The same approach was used to determine the top 1% most highly cited publications by main fields of science.

publication journals that is based on a completely different set of publications than the set included in the analysis (i.e. publications from 2011).

The altmetric indicators obtained through Altmetric.com are the followings:

- *Facebook walls*: number of times a publication has been mentioned on a wall in Facebook.
- *Blogs*: number of times a publication has been mentioned in blogs
- *Twitter*: number of Twitter users that have tweeted (or re-tweeted) a publication.
- *Google+*: number of Google+ users that have mentioned the publication.
- *News outlets*: number of mentions of scholarly articles in magazines and news outlets[9].
- *Total Altmetrics*[10]: this indicator has been calculated by summing all the values from the previous altmetric scores. It is important to highlight that when a publication does not have any score in the altmetric indicators, we consider this to be a '0' and not a missing value. The reason for this choice is that although some publications may have errors in their identification through altmetrics sources, the fact that a publication is not mentioned at all can be considered to be a 0, as it has no altmetrics (as when a publication is not cited at all and it gets a citation score of 0)[11].This compound indicators is calculated only for exploratory reasons and in order to simplify the analysis and reduce the number tables and graphs, but this does not mean that we propose it as an indicator in itself (in fact, our results show that different dimension of altmetric indicators could be suggested).

---

[7] The reason to select the 1 % of the publications based on the ncs is the strong differences in impact that we can find in a multidisciplinary dataset as it is the case for the one studied here, as well as the important sub-disciplinary differences within the major fields of science considered in this study. Thus we believe that we have selected the most genuine top highly cited publications within our dataset.

[8] In case of ties we sorted by the fact that the publication was also among the top 10% within its subject category, then descending by publication month (thus giving priority to most recent publication that would have a better altmetric coverage) and finally by UT descending (thus expecting that we select also the most recent publications). The main idea thus was to give priority to genuinely highly cited publications as well as the most recent ones.

[9] Altmetric.com collects online mentions of scholarly papers from reports published in mainstream news outlets and magazines. They do this by tracking a manually-curated list of RSS feeds from news websites. They add each news source individually, and also try to get cover outlets in non-English-speaking countries.

[10] This is not among the indicators provided by altmetric.com , it is calculated by ourselves.

[11] Somehow, mistakes in the detection of altmetric mentions can be equated to the mistakes in citation linkages that can still be found in citation databases (Schmidt, 2012)

For the disciplinary analysis we have considered the classification of Science in 5 major fields presented in the Leiden Ranking (Waltman, et. al., 2012 - http://www.leidenranking.com/methodology/fields) which basically consists on the aggregation of WoS Subject Categories in five main disciplinary fields. In our case, for simplicity we haven't fractionalized publications or citations and altmetric indicators, so a publication can be linked and fully counted in different fields. Finally, statistical analysis have been performed using IBM SPSS 21 and Matlab R2012b.

### 3. Results

In this chapter we present the main results of our study. The results are presented in 3 different sections. In the first section we present the results regarding the presence and frequency of altmetrics across scientific publications. The second section studies the relationships among all these altmetric and bibliometric indicators and the third section pays special attention to the relationships between altmetrics and citations.

*3.1 Presence of publications with altmetrics and relationship with citations and journal indicators*

In this section we study the presence of altmetrics across the 2011 July-onwards population of publications. As it can be seen in tables 1 and 2, the final number of publications included in the analysis amounts to 500,229, of which 75,569 (15%) have at least one altmetric score.

**Table 1. General distribution of publications with altmetrics**

|  | Docs. | % within altmetrics | %pubs |
|---|---|---|---|
| Total altmetrics | 75569 | 100% | 15.1% |
| Facebook walls | 12386 | 16.4% | 2.5% |
| Blogs | 9444 | 12.5% | 1.9% |
| Twitter | 66591 | 88.1% | 13.3% |
| Google+ | 3021 | 4.0% | 0.6% |
| News outlets | 2331 | 3.1% | 0.5% |

The source that provides more altmetrics scores is Twitter with 13% of all the publications that get some Twitter mentions, followed at an important distance by Facebook (2.5%), mentions in blogs (1.9%), Google+ accounts mentions (0.6%) and finally mentions in news outlets (0.5%). This is in line with Thelwall, et.al. (2013) in which they found that except for Twitter, the coverage of all altmetrics sources for PubMed articles were very low (substantially below 20%).

Table 2 presents a more detailed analysis of the distribution of altmetrics and citations across fields. As we can see the fields with the highest share of publications with altmetric scores are the 'Biomedical and health sciences' and the 'Social sciences and humanities' with more than 22% of the publications having at least one altmetric score. 'Life and earth sciences' present altmetrics in less than 20% of the publications, and the 'Natural sciences and engineering' and 'Mathematics and computer science' have less than 10% of their publications with altmetrics.

**Table 2. General presence of altmetrics across major fields of science**

| Fields | p | tcs | mcs | Facebook | Blogs | Twitter | Google+ | News | Total Altmetrics | Alt/pubs | Pubs. With Alt | % pubs with alt |
|---|---|---|---|---|---|---|---|---|---|---|---|---|
| Biomedical and health sciences | 217115 | 451111 | 2.1 | 15821 | 7758 | 151454 | 3530 | 1809 | 180372 | 0.83 | 49575 | 22.83% |
| Life and earth sciences | 100286 | 163922 | 1.6 | 4632 | 5236 | 57167 | 2066 | 1826 | 70927 | 0.71 | 15989 | 15.94% |
| Mathematics and computer science | 51730 | 33439 | 0.6 | 841 | 858 | 12989 | 672 | 256 | 15616 | 0.30 | 2788 | 5.39% |
| Natural sciences and engineering | 172094 | 264482 | 1.5 | 2428 | 3993 | 37116 | 1829 | 1088 | 46454 | 0.27 | 15456 | 8.98% |
| Social sciences and humanities | 45445 | 39454 | 0.9 | 2295 | 2931 | 39758 | 1705 | 682 | 47371 | 1.04 | 10226 | 22.50% |
| Total | 500229 | 796321 | 1.6 | 19956 | 14326 | 209228 | 5813 | 3476 | 252799 | 0.51 | 75569 | 15.11% |

Regarding the citation density and the altmetrics density (i.e. the average number of citations or altmetrics per publication), we can see how the highest altmetrics density is for publications published in the Social sciences and humanities (1.04), followed by the Biomedical and health sciences (0.83) and Life and earth sciences (0.71).

From another perspective, if we focus on the differences between the citation density (i.e. the mcs indicator) and the altmetrics density, we can see how the field of Social sciences and humanities has actually a slightly higher density of altmetrics per paper than citations (1.04 vs. 0.9), while for the other fields altmetrics have always a lower density compared to citations.

*Factor analysis of bibliometric and altmetric indicators*

In this section we put the accent on the analysis of the factor analysis and correlations among the different indicators. In table 3 we present a factor analysis of the previous indicators.

**Table 3 . Factor analysis (loadings > 0.5 highlighted in bold, and > 0.2 in italics) – Rotated component matrix – Varimax rotation**

|  | Component | | | |
|---|---|---|---|---|
|  | 1 | 2 | 3 | 4 |
| jcs_0510_12 | **,920** | ,062 | ,136 | ,083 |
| jfis | **,914** | ,047 | ,154 | ,112 |
| jfis_0510_12 | **,904** | ,041 | ,119 | ,110 |
| jcs | **,903** | ,059 | ,160 | ,070 |
| Total altmetrics | ,088 | **,933** | ,047 | *,249* |
| Twitter | ,087 | **,895** | ,043 | *,215* |
| Google+ | -,012 | **,700** | -,017 | ,106 |
| Facebook walls | ,041 | **,697** | ,043 | -,025 |
| ncs | ,198 | ,042 | **,941** | ,095 |
| cs | *,231* | ,040 | **,938** | ,056 |
| News outlets | ,126 | ,103 | ,046 | **,882** |
| blogs | ,142 | *,321* | ,109 | **,773** |

Extraction Method: Principal Component Analysis.
Rotation Method: Varimax with Kaiser Normalization.
Rotation converged in 5 iterations.

Factor analysis shows four main components or dimensions. 80% of the total variance is explained by this factor analysis (based on Principal Component Analysis).The first and the third dimensions are related to bibliometric indicators. In fact, the first dimension is related to journal-based indicators and the third dimension is related to the observed impact of the publication (i.e. their actual number of citations and normalized citations). This separation of journal-based indicators and article-based indicators have been already found in previous studies (cf. Costas, van Leeuwen, & Bordons, 2010) and basically is in line with the conceptual delineation of indicators that measure the impact of the publication venue (e.g. journals) and indicators that measure the direct impact of publications.

From another perspective, altmetric indicators are also split in two main dimensions (components 2 and 4). Dimension 2 shows the correlations of the total altmetrics indicator with Twitter, Facebook and Google+. The second one is composed of blogs and news mentions. Thus, mentions in blogs and news outlets seem to represent another different type of impact 'flavor' as compared to Tweets, Facebook and Google+ mentions. In this regards, we obtain different result as compared to the results of Priem et. al. (2012b) where they found that Facebook was actually not correlated with Tweets while blogs where correlated with Twitter. In order to further test this point, in Appendix II we present the Factor Analysis for the all the publications from Altmetric.com with a DOI (i.e. 1,380,143 publications). Results are again quite consistent with our previous finding: blogs and news conform to a different dimension compared to Tweets, Facebook and Google+ mentions. Differences in populations and methods of counting altmetrics could be among the explanation between the two studies.

*Correlations among bibliometric and altmetric indicators*

In this section we study the correlations between altmetrics and bibliometrics in our population of 500,229 publications. Table 4 presents the Pearson's correlation analysis of the rank values of the indicators considered[12], thus we are providing the Spearman correlation of the original variables. Confidence intervals at 95% are presented between brackets obtained through the bootstrapping technique implemented in SPSS (based on 1000 re-samplings).

**Table 4. Correlation analysis of the rank values of the main citation and altmetrics variables (loadings > 0.250 are highlighted in bold characters and >0.150 in italics)**

|  | Rank of cs | Rank of ncs | Rank of jfis | Rank of jfis_0510_12 | Rank of jcs | Rank of jcs_0510_12 | Rank of Facebook | Rank of blogs | Rank of Twitter | Rank of Google+ | Rank of news | Rank of Total altmetrics |
|---|---|---|---|---|---|---|---|---|---|---|---|---|
| Rank of Cs | 1 (1-1) | **0.97 (0.969-0.97)** | **0.39 (0.388-0.392)** | **0.527 (0.525-0.529)** | **0.34 (0.337-0.342)** | **0.478 (0.475-0.48)** | 0.099 (0.096-0.101) | 0.126 (0.123-0.129) | *0.167 (0.164-0.17)* | 0.06 (0.057-0.063) | 0.076 (0.073-0.079) | *0.184 (0.181-0.186)* |
| Rank of Ncs |  | 1 (1-1) | **0.393 (0.391-0.396)** | **0.442 (0.439-0.444)** | **0.346 (0.344-0.349)** | **0.407 (0.405-0.409)** | 0.086 (0.083-0.089) | 0.112 (0.11-0.115) | *0.141 (0.138-0.144)* | 0.053 (0.05-0.056) | 0.069 (0.066-0.072) | *0.156 (0.153-0.159)* |
| Rank of Jfis |  |  | 1 (1-1) | **0.738 (0.737-0.739)** | **0.897 (0.896-0.898)** | **0.703 (0.702-0.705)** | 0.088 (0.086-0.091) | 0.125 (0.122-0.127) | *0.172 (0.169-0.174)* | 0.061 (0.058-0.064) | 0.075 (0.072-0.077) | *0.187 (0.184-0.189)* |
| Rank of jfis_0510_12 |  |  |  | 1 (1-1) | **0.644 (0.642-0.646)** | **0.907 (0.906-0.908)** | 0.11 (0.107-0.113) | 0.132 (0.129-0.134) | *0.227 (0.225-0.23)* | 0.069 (0.067-0.072) | 0.074 (0.071-0.077) | **0.244 (0.242-0.247)** |
| Rank of Jcs |  |  |  |  | 1 (1-1) | **0.736 (0.735-0.738)** | 0.086 (0.083-0.089) | 0.124 (0.121-0.126) | *0.169 (0.166-0.171)* | 0.063 (0.06-0.066) | 0.074 (0.071-0.077) | *0.183 (0.18-0.186)* |
| Rank of jcs_0510_12 |  |  |  |  |  | 1 (1-1) | 0.108 (0.105-0.11) | 0.13 (0.127-0.132) | *0.223 (0.221-0.226)* | 0.067 (0.064-0.07) | 0.072 (0.07-0.075) | *0.24 (0.237-0.242)* |
| Rank of Facebook |  |  |  |  |  |  | 1 (1-1) | *0.177 (0.17-0.184)* | **0.256 (0.251-0.26)** | *0.15 (0.141-0.159)* | *0.145 (0.137-0.155)* | **0.394 (0.391-0.397)** |
| Rank of Blogs |  |  |  |  |  |  |  | 1 (1-1) | *0.206 (0.201-0.21)* | *0.198 (0.187-0.208)* | **0.267 (0.256-0.278)** | **0.343 (0.34-0.347)** |
| Rank of Twitter |  |  |  |  |  |  |  |  | 1 (1-1) | 0.148 (0.143-0.153) | 0.137 (0.132-0.142) | **0.935 (0.934-0.936)** |
| Rank of Google+ |  |  |  |  |  |  |  |  |  | 1 (1-1) | *0.197 (0.182-0.212)* | *0.198 (0.194-0.201)* |
| Rank of News |  |  |  |  |  |  |  |  |  |  | 1 (1-1) | *0.175 (0.172-0.179)* |
| Rank of Total altmetrics |  |  |  |  |  |  |  |  |  |  |  | 1 (1-1) |

Table 4 basically supports the results previously presented in the factor analysis. In the first place, bibliometric indicators correlate the most among them and the same holds for altmetric indicators.

---

[12] Given some technical limitation of SPSS to calculate the Spearman correlation on a sample as big as ours, we have followed the recommendation provided by IBM of ranking the data (from largest to smallest and assigning the highest value to the ties) and calculating Pearson's correlation (thus, virtually reproducing a Spearman correlation). For more details check http://www-01.ibm.com/support/docview.wss?uid=swg21476714

Within the bibliometric indicators, article-based indicators have a stronger correlation among them (i.e. cs and ncs) and with journal indicators, which correlate quite highly among themselves (particularly the pairs of JCS and JFIS of the same period). In any case, these results corroborate that bibliometric indicators correlate better among themselves than with altmetrics indicators and that there are two conceptual dimensions of indicators: article-based and journal impact indicators.

On the altmetrics side, the compound indicator total altmetrics correlates mostly with Twitter, which makes sense as Twitter dominates in this indicator; and both indicators correlate only moderately with citations and journal indicators. The highest correlations with the bibliometric indicators are with the journal indicators that refer to the previous period (i.e. 2005-2010). The other indicators have quite low correlations among them, although it is still worthwhile to mention the correlation between blogs and news outlets and between Twitter and Facebook.

All in all, the two individual altmetric indicators that have more relation with citations and journal indicators are Twitter and blog mentions, all the other metrics have negligible correlations with citation based impact indicators.

### 3.2. Comparison between altmetrics and citations

Previous results are quite consistent and indicate a positive but weak correlation between altmetrics and citations. In this section, we study more directly this relationship between altmetric indicators and the impact of the publications. The focus is to analyze whether altmetrics have some predictive power on citations and particularly on identifying highly cited publications. Table 5 presents the relationship between the number of altmetrics (i.e. total altmetrics) and the number of citations (both non-normalized and field normalized) and with the impact of journals (JCS).

**Table 5. Average number of citations and JCS vales of publications with different numbers of altmetric scores (between brackets 95% confidence intervals based on bootstrapping)**

| N. Altmetrics | N. publications | Mean citations | Mean normalized citations | Mean JCS |
|---|---|---|---|---|
| 0 | 424660 | 1.28 (1.27-1.28) | 0.6976 (0.6937-0.7018) | 1.8485 (1.8426-1.8548) |
| 1 | 43078 | 2.36 (2.27-2.51) | 1.0239 (0.9938-1.0698) | 3.0506 (3.0195-3.0803) |
| 2 | 13585 | 2.96 (2.87-3.05) | 1.2584 (1.2179-1.2984) | 3.7527 (3.6695-3.8294) |
| 3 | 5920 | 3.79 (3.6-3.99) | 1.5559 (1.4808-1.6359) | 4.4605 (4.3045-4.6412) |
| 4 | 3238 | 4.23 (3.97-4.52) | 1.7205 (1.6122-1.8441) | 4.7279 (4.5105-4.9574) |
| 5 or more | 9748 | 7.85 (7.49-8.21) | 3.1642 (3.0305-3.2948) | 7.9147 (7.7258-8.1143) |
| Total | 500229 | 1.59 (1.58-1.61) | 0.8058 (0.7999-0.8115) | 2.1715 (2.1636-2.1795) |

As shown in table 5, the average number of citations per publication (both normalized and non-normalized) and the average JCS per publication increase with the number of total altmetrics. It is a quite strong pattern, particularly between the value 0 and the rest, where we can see how the field normalized impact moves from below 1 (i.e. below the international level determined by the value of 1) to above 1 with just one altmetric score.

Figures 2 and 3 graphically show the previous relationships. The figures also display 95% confidence intervals (for larger numbers of altmetrics, the confidence intervals are wider due to the relatively smaller numbers of publications with these numbers of altmetrics). As it can be seen there is a clear positive relation between the number of altmetrics and the average citation impact and JCS of the publications, in a way that publications with more altmetrics also tend to have more citations. This has been found in previous studies as well (Waltman & Costas, 2013; Zahedi, Costas & Wouters, 2013). However, the interesting issue now is to qualify this relationship between citations and altmetrics. This is performed in the following section using the methodology of precision-recall analysis previously developed by Waltman & Costas (2013).

**Figure 2. Relations between the number of total altmetrics and the number of citations**

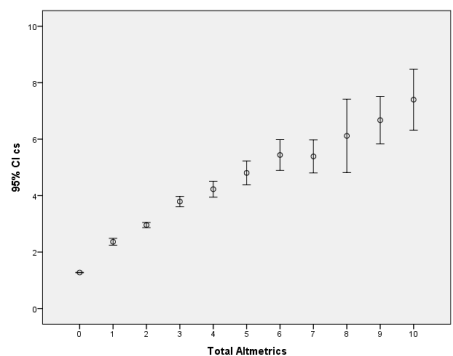

**Figure 3. Relations between the number of total altmetrics and the JCS**

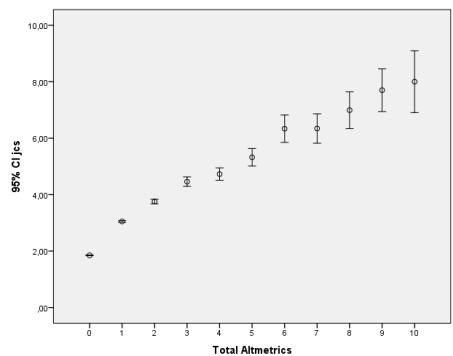

*Precision-recall analysis*

All our previous results point towards the idea that citations and altmetrics are related and that this relationship is strong when we look at the aggregation of publications as for example demonstrated by Figures 2 and 3. However, the correlations at the paper level are rather weak and particularly they are much weaker than the correlation between journal indicators and citations. These low correlations at the paper level are not a surprise if we take into account that around 85% of all the publications in our dataset do not have any altmetric score, therefore the usefulness of correlation analysis is lower here. Another approach to study the relationship between citations and altmetrics can be the precision and recall analysis developed in Waltman & Costas (2013). This analysis allows the test of altmetrics as tools that can help their users to identify (and thus 'filter') highly cited publications, and particularly we can test if they do it better than for example journal citation scores.

We calculate precision-recall figures in order to identify the top 1% most highly cited publications in our dataset (sorted by their ncs value). For a given selection of publications, precision is defined as the number of highly cited publications in the selection divided by the total number of publications in the selection. Recall is defined as the number of highly cited publications in the selection divided by the total number of highly cited publications. We focus on all publications that are among the 1% most cited (based on the ncs indicator) and we test them for the JCS_0510_12 (thus using a journal indicator that is independent form our sample of publications[13]) and for the total altmetrics indicator. Figure 4 presents the result for the whole set of publications.

The interpretation of the precision-recall curves is as follows: take the curve obtained for total altmetrics (green line) , this curve indicates that a recall of 0.10 (10%) corresponds with a precision of around 0.25 (25%), this meaning that if we want 25% of the publications in our selection to belong to the top 1% most highly cited publications, our selection can manage to include only 10% of all top 1% most cited publications.

**Figure 4. Precision-recall curves for JCS (blue line) and total altmetrics (green line) for identifying 1% most highly cited publications – all publications**

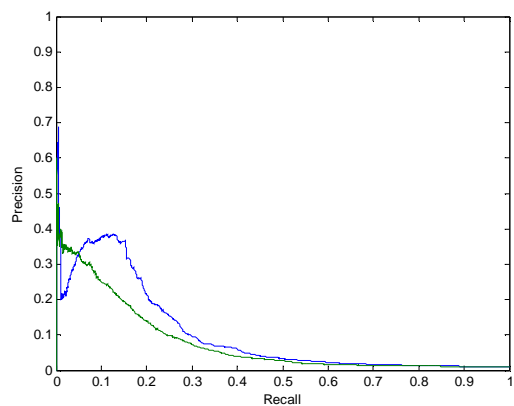

Figure 4 shows that JCS performs in general better than altmetrics in identifying the top 1% most highly cited publications within our dataset. Although with values of recall around ~0.05 (i.e. ~5%) altmetrics actually outperform the JCS indicator. This means that in those cases where recall is not really an issue (e.g. scholars with no much time to read papers and just interested in reading the most cited one) altmetrics could indeed play a role as an alternative approach to find highly cited papers.

---

[13] For example, it could be argued that this information is available at the moment of publication of the paper and we don't need to wait for the citations to be gathered.

Complementary to Figure 4 we have also checked the precision-recall figures of the individual altmetric indicators (appendix III). In general, most of the altmetric indicator do not change the observations of the general pattern, although it is remarkable that blogs show an interesting pattern when it comes to precision, as the values of precision are higher than those of the JCS indicator for the lower levels of recall. It is remarkable that blog mentions (and also news mentions) present higher levels of precision in identifying highly cited publications. In order to explore in more detail this higher precision of blogs for filtering highly cited publications in Figure 5 we present the precision-recall curves for blogs and twitter mentions. In this case, it is noticeable that blogs have a higher precision in identifying top papers than Twitter, but it is also remarkable that with higher levels of recall Twitter does not really improve the capacity of blog mentions of filtering highly cited publications (as their precision-recall curves go quite parallel through the whole recall spectrum after values of ~0.1). This suggests that among all altmetric scores, blogs have a stronger potential in identifying highly cited publications than for example tweets (in line with the suggestions by Shema, et.al., 2013).

**Figure 5. Precision-recall curves for Twitter (blue line) and blogs mentions (green line) for identifying 1% most highly cited publications – all publications**

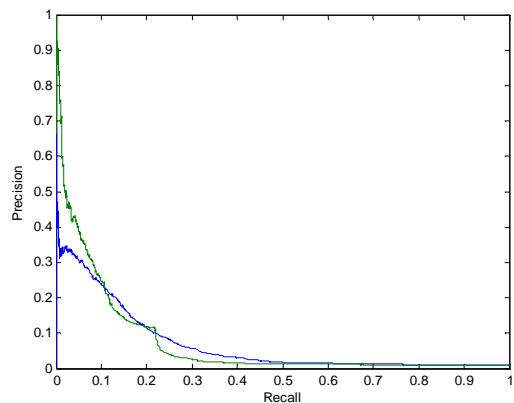

In figure 6 we present the same precision-recall curves for the 5 major fields of science. The methodology and interpretation of the indicators is the same as in Figure 4. It is important to keep in mind that the top 1% most highly cited publications have been calculated individually for every field of science (based on their ncs indicator), thus adapting the selection of top 1% of highly cited publications to every discipline.

**Figure 6. Precision-recall curves for JCS (blue line) and total altmetrics (green line) for identifying 1% most highly cited publications**

**Biomedical and health sciences**

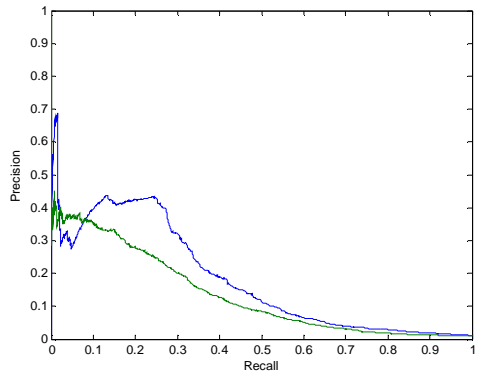

**Life and earth sciences**

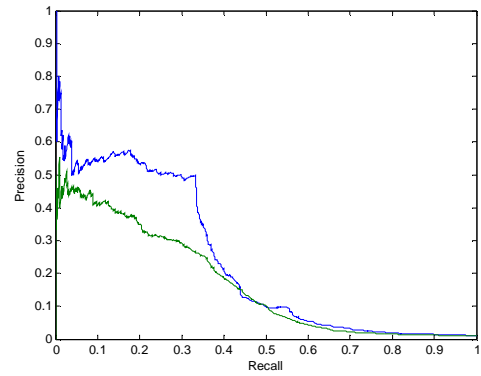

**Mathematics and computer science**

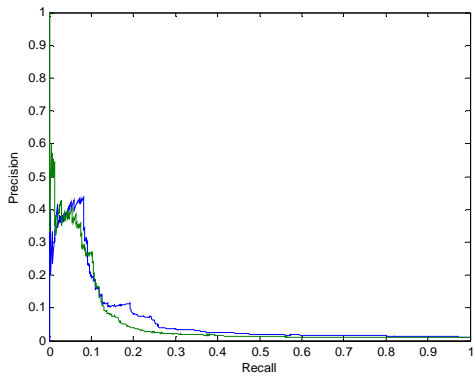

**Natural Sciences and engineering**

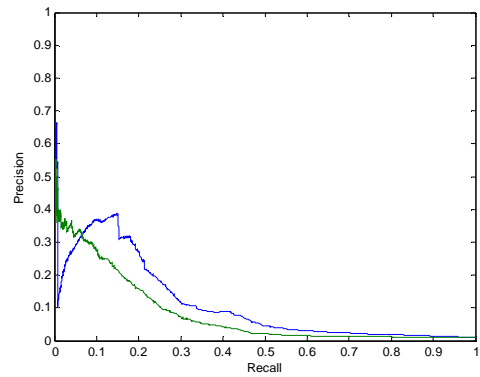

**Social sciences and humanities**

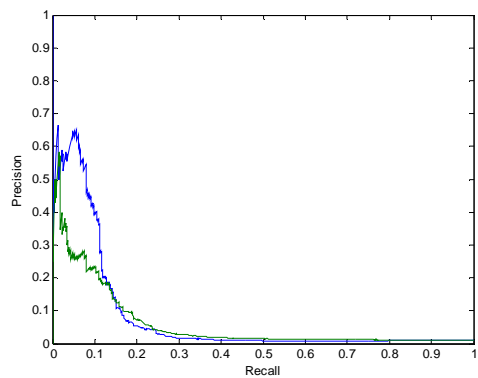

The analysis of figure 6 shows some interesting patterns. In the first place, the 'Biomedical and health sciences' and the 'Natural sciences and engineering' resemble the general pattern of general levels of precision and recall for the JCS scores. These two are the biggest disciplines in our dataset as observed in table 2, thus explaining the strong similarity with the general pattern.

The 'Life and earth sciences' also resemble the general pattern although in this case the advantage of altmetrics in precision within the lowest levels of recall is not observed. In this discipline JCS outperforms altmetrics along the whole spectrum of the precision-recall lines.

The 'Social sciences and humanities' present an interesting pattern. Although in general JCS scores outperform altmetrics scores for the lowest levels of recall (<0.10), from that point onwards both measures tend to merge, although also in both cases with quite low levels of precision.

Finally, the discipline of 'Mathematics and computer science' presents the most distinct pattern. In this case, in addition to the general low levels of precision and recall, we also have that none of them have a real advantage over the other in filtering highly cited publications. They mostly overlap along of the precision-recall lines, thus indicating a bad filtering capacity of both JCS and altmetrics regarding top publications in this field.

### 3.3. Tight analysis

All our previous analyses suggest that altmetrics have only a limited correlation with citations and a relatively lower value for extensively identifying top publications (i.e. with relatively high levels of recall), at least not better than journal indicators. Partly, this lower recall of altmetrics can be explained by the fact that in our population 55% of the top 1% most highly cited publications have no altmetrics at all. In a way, this suggests a weakness in the capacity of altmetrics to identify highly cited publications in a full universe of publications (e.g. considering the whole WoS set of publications). However, one could argue that the altmetric community (e.g. bloggers, Twitter users, etc.) is perhaps small and therefore it is not possible to reach all the scientific publications and filter them. In addition, the mission of these tools is not really to filter the 'best' publications (at least in terms of highly cited publications), therefore it is not of such importance if they are not able to filter and select all highly cited publications (or most of them).

In any case, given this limited reach of altmetrics, we could wonder what would happen if we would limit our population to only those publications that have at least one altmetric score, thus creating an 'altmetrics-driven' universe of publications, where all publications have had some altmetric filtering. In other words, we wonder if the analysis of the relationship between altmetrics and citations would improve if we focus only on those publications that have been at some point picked up by the social media community (i.e. twitters, bloggers, etc.). We term this analysis as a 'tight analysis' as it is limited only to the 73,711 publications from July 2011 onwards that have at least 1 altmetric score[14].

In Appendix IV we present the main correlations of this tight analysis and in Appendix V the main precision-recall figures (general and for the main fields). The main conclusion based on those results is that there are only marginal improvements in the relationship between altmetrics and impact indicators. Perhaps one of the highest improvements is for blogs with JCS indicators (Figure 7) where we can see how the recall values increase, while precision stays higher for blog mentions than for JCS. In any case, in general terms the whole picture that comes out of the 'tight analysis' is that the same results as previously presented are still observed.

---

[14] Notice that this 'tight analysis' also has the advantage that it suffers less from the limitation of the lack of identification of papers with an early 'online first' publication, given the situation that now all publications have altmetrics and no publications with missing altmetrics could influence the analysis.

**Figure 7. Precision-recall curves for JCS (blue line) and blogs (green line) for identifying 1% most highly cited publications-tight analysis**

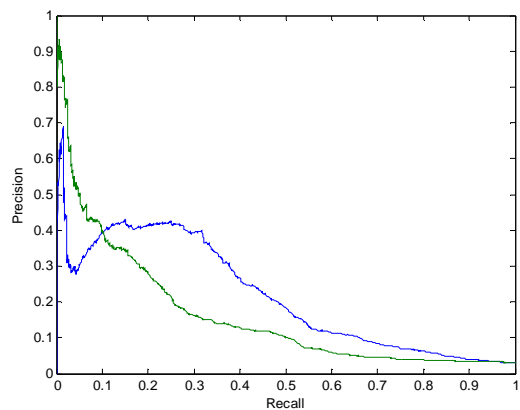

**Discussion**

*Limitations of the study*

In this paper we have performed an extensive and multidisciplinary analysis of the presence of altmetric data (excluding Mendeley) across scientific fields and we particularly focus on their correlations with citations. In the first place, it is important to acknowledge and contextualize some of the main limitations involved in this study that go beyond the regular limitations regarding altmetrics studies (particularly regarding data quality). The first limitation is related to the coverage of Altmetric.com. Due to the admitted restriction of Altmetric.com of being more robust for publications published from July 2011 onwards, we couldn't work with a full publication year, therefore problems related with the proper identification of the publication month of the publications needed to be observed (e.g. not all publications consistently indicate their publication month). Besides, the publication month of an article is not necessarily always indicative of when the paper appear to the public (e.g. they could appear before through "online first" versions). For this reason some altmetric data for the publications of the first months could have been lost. In any case, the broad scope of the analysis and the number of publications from the first months of the study ensure the robustness of our analysis. Also, the corroboration of the main results based on the 'tight analysis' supports the validity of our study.

All in all, we consider that we have combined in a quite balanced way both bibliometric and altmetric indicators. Future analysis (e.g. the analysis of publications from 2012) should corroborate (or perhaps discuss) the results presented here. For the time being however, we can consider the results presented here as a valid first exploratory approach and the results are generalizable insofar as the limitations do not invalidate the main conclusions.

*Summary of the results and main conclusions*

- *General analysis*

Our study indicates that around 15% of the publications from 2011 (July onwards) have any altmetric measures, although the percentage of publications with altmetrics scores is increasing for the most recent years. For example, for the year 2012 the share of publications with any altmetric measure is above 20%. Based on this we can conclude that altmetrics are only valid and valuable for the most recent publications. This higher presence of Twitter metrics for the most recent years has been also observed by Haustein et. al. (2013b) with values similar to the ones presented in this paper.

However, even considering this increasing presence of altmetrics for the most recent years, it is possible to discuss what the role of altmetrics could be for research assessments, particularly if they can be seen as potential replacements (or alternatives) to citations. In a way, the lower presence of publications with altmetrics (even for the most recent years) could challenge the reliability of any development of indicators based on altmetrics. To illustrate this point, in 2012 around 24% of all the publications with a DOI presented some altmetric scores, while in the same year 26% of the publications had already received at least one citation in the same year. This means that even for this recent year, the number of publications with citations outperforms the number of publications with altmetrics. However, more important is that over the next months and years we can expect an increase in the number of citations for 2012 publications (i.e. this 26% of publications with citations will naturally increase over time) thus increasing the information on citation impact for scientific publications (and therefore increasing the reliability of indicators based on citations), while the number of publications from 2012 with altmetrics scores is not expected to increase significantly over time as altmetrics are a very immediate and fast type of impact events (cf. Haustein et al, 2013b).

Based on these results we can also argue that even considering the increase in the number of publications with altmetric impact for the most recent years, if the number of altmetrics does not really increase (i.e. not much higher than 20-30%), the potential value of altmetrics would be limited by the lack of information for most of the publications.

One of the most interest results of this paper is that altmetric counts do not always present a better filtering of highly cited publications than journal citation scores. We observed that altmetrics scores are able to filter top publications with higher levels of precision than journal scores (particularly blog mentions) but they have a lower recall than JCS in identifying highly cited publications. A practical interpretation of these results can be as follows: a researcher with very little time to read can better filter highly cited top publications based on altmetrics (i.e. selecting just a few highly cited publications based on their high altmetrics counts), while for example a library interested in selecting as many top cited publications as possible would still be better served by the JCS scores. Overall, it can be claimed that our results help to qualify the idea of altmetrics as potential filtering tools of relevant publications (at least if we equate the idea of relevance to those publications that have a high scientific impact) in the sense that altmetrics can indeed help to filter them but with a limited capacity in reaching all of them. A similar situation was observed for F1000 recommendations by Waltman & Costas (2013) and somehow here we can extend this observation to other altmetrics measures.

From another perspective, it is remarkable that altmetrics coming from mentions in blogs and news outlets have a relatively stronger correlation with citations compared to the other altmetrics measures. The fact that these metrics go together in our factor analysis suggests the idea of scientific blogs as a new genre of scientific outputs on their own that share characteristics with other means of the scientific discourse (Shema et. al., 2013). For example the stronger correlations between blog and news mentions with journal indicators and citations supports the observation by Shema et. al. (2013) and Groth & Gurney (2010) that bloggers regularly cite well-known, high-impact journal publications. The moderate correlation between blogs with twitter scores also supports the claim by Shema et. al. (2013) that bloggers are information disseminators in more than one social medium.

All in all, our study also confirms the results from previous studies (Priem, Piwowar, & Hemminger, 2012; Bar-Ilan, 2012; Waltman & Costas, 2013; Zahedi, Costas & Wouters, 2013; Thelwall et. al., 2013; Haustein et. al., 2013b) that the relationships between altmetrics and citations and journal impact scores are positive but only moderate. The fact that we find a relatively weak correlation between citations and altmetrics both in the general analysis and in the tight analysis supports the idea that altmetrics do not really reflect citation impact, thus supporting previous claims by Haustein et. al. (2013b) that citations and altmetrics very likely measure different types of impact. Therefore, the potential of altmetrics as a replacement of citations (also considering the lower presence and density of altmetrics among scientific publications) as measures of scientific impact seems quite improbable. However, the question about the potential complementarity of altmetrics as a source of evidence of other types impact not captured by citations is still open and this point needs to be explored in future research.

- *Disciplinary analysis*

The analysis of the different major fields of science also displays interesting results. In the first place there are differences by fields in terms of the share of publications with any altmetric score. Publications from the social sciences and humanities exhibit a higher altmetric activity (Zahedi, Costas & Wouters, 2013) and their altmetrics density is quite similar to their citation density. This finding suggests that altmetrics scores could have an interesting added value for the analysis of humanities and social sciences, fields that traditionally are not well represented by traditional citation analysis. This finding supports the idea that altmetrics measures could be related with the more cultural or social aspects of scientific work, with their presence among the social and humanistic fields being its main exponent. Further research should delve into these ideas to confirm this advantage of altmetrics to support the analysis of these fields and their relationship with the more societal 'impact' of scientific outputs.

Our analysis also shows that the medical and life sciences present a comparatively high presence of publications with some altmetric scores (around 19%) which is in line with the findings of Haustein et. al. (2013b) in their analysis of PubMed publications. The altmetrics density is nevertheless considerably lower compared to the citation density, indicating that citation indicators may still play a more prominent and informative role in this area.

Finally, Mathematics and computer science and Natural sciences and engineering sciences are the fields with the lowest presence and density of altmetrics per paper. Given these results we can argue that for these fields altmetrics have a lower chance to become an alternative or complementary to citations, unless their frequency and density increase in the next years.

**Final conclusions**

This study poses several relevant conclusions in the research of social media indicators and altmetrics. In the first place, our results confirm that altmetrics are only valid for the most recent publications and that their presence is increasing over time. Thus, depending on how the incorporation of these social activities among scholars and the general public evolves, they could improve their role and validity for the complementary analysis and evaluation of scientific publications.

From this study, it is also possible to conclude that the presence and density of social media altmetrics is still low among scientific publications, thus challenging the reliability of indicators based on them. In addition, the fact that they exhibit only weak correlations with citations suggests that the potential of altmetrics as a replacement of the more traditional citation analysis is not very strong. However, they could actually represent an interesting relevant complement to citations, particularly in order to inform other types of impact (e.g. societal or cultural impact) and specially in those fields where they have a higher presence, mostly the humanities and social sciences. In this sense, more research is necessary in order to determine and validate these other potential types of impact, probably combining not only quantitative analysis as in this paper but also other more qualitative studies as already suggested by other studies (Thelwall, et.al., 2013;Haustein et. al., 2013b; Zahedi, Costas & Wouters, 2013).

**Acknowledgements**

The authors of this paper wish to thank Euan Adie from Altmetric.com for his help and support in working with Altmetric.com data, also the support by Erik van Wijk from CWTS in managing altmetric data. Finally, the authors acknowledge the comments and suggestions by Ludo Waltman from CWTS on early versions of this paper.

**Appendix I. Evolution of publications and altmetrics from July 2011 onwards**

| Pub. month in 2011 | p | p with 'altmetrics' | % altmetrics |
|---|---|---|---|
| 7 | 79521 | 8468 | 10.6% |
| 8 | 78024 | 9847 | 12.6% |
| 9 | 87228 | 12385 | 14.2% |
| 10 | 84797 | 13592 | 16.0% |
| 11 | 80499 | 14703 | 18.3% |
| 12 | 90160 | 16574 | 18.4% |
| total | 500229 | 75569 | 15.1% |

**Appendix II. Factor analysis of all altmetric indicators of all publications with a DOI from Altmetric.com**

**Rotated Component Matrix[a]**

| | Component | |
|---|---|---|
| | 1 | 2 |
| Total Altmetrics | **,946** | ,189 |
| Twitter | **,915** | ,141 |
| Facebook walls | **,738** | ,095 |
| Google+ | **,581** | ,177 |
| News outlets | ,111 | **,820** |
| Blogs | ,193 | **,773** |

Extraction Method: Principal Component Analysis.
Rotation Method: Varimax with Kaiser Normalization.
Rotation converged in 3 iterations.
67% of the total variance explained.

**Appendix III. Precision-recall analysis of individual altmetric indicators (green lines) vs JCS (blue lines): extended analysis**

**Twitter vs. JCS**

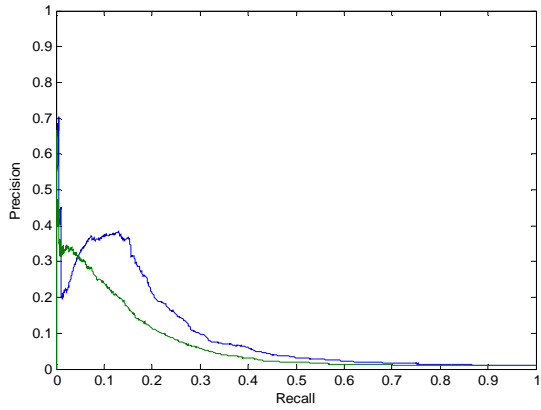

**Facebook vs. JCS**

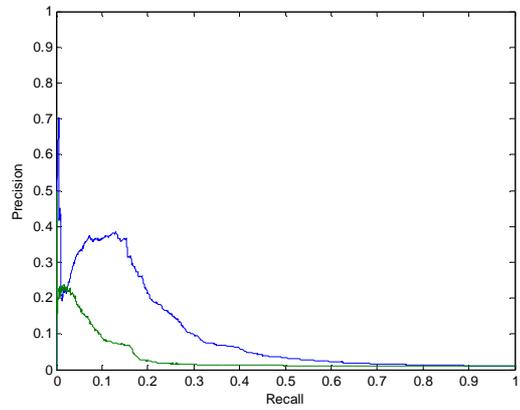

**Blogs vs. JCS**

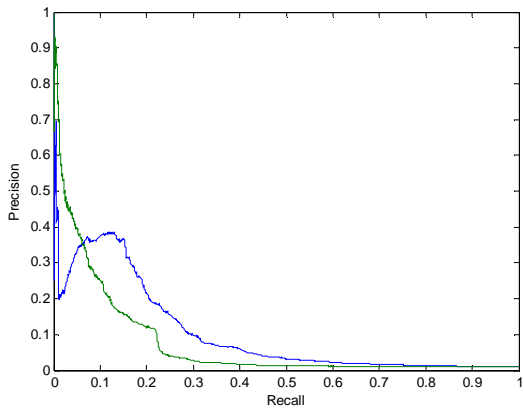

**Google+ vs. JCS**

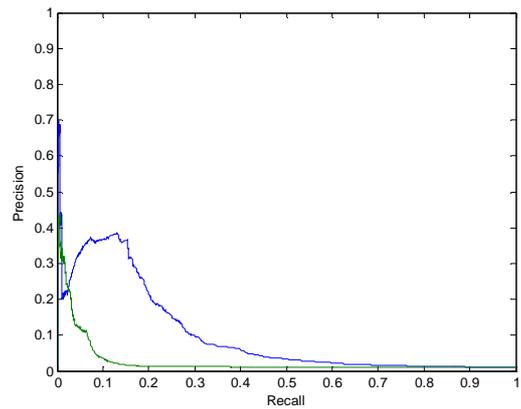

**News vs. JCS**

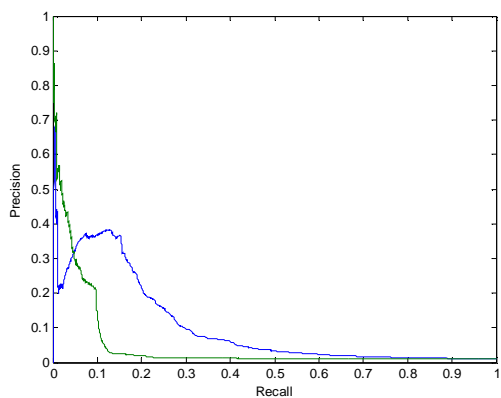

**Appendix IV. Correlation analysis – 'Tight analysis' (loadings > 0.25 are highlighted in bold characters and >0.15 in italics)**

| | Rank of cs | Rank of ncs | Rank of jfis | Rank of jfis_0510_12 | Rank of jcs | Rank of jcs_0510_12 | Rank of facebook | Rank of blogs | Rank of twitter | Rank of gplus | Rank of news | Rank of total altmetrics |
|---|---|---|---|---|---|---|---|---|---|---|---|---|
| Rank of cs | 1 (1-1) | **0.951 (0.95-0.952)** | **0.455 (0.449-0.461)** | **0.57 (0.565-0.574)** | **0.404 (0.397-0.41)** | **0.51 (0.505-0.515)** | 0.097 (0.09-0.104) | *0.199 (0.191-0.206)* | *0.161 (0.154-0.168)* | 0.077 (0.069-0.084) | 0.134 (0.127-0.142) | *0.195 (0.188-0.202)* |
| Rank of ncs | | 1 (1-1) | **0.451 (0.445-0.457)** | **0.442 (0.437-0.448)** | **0.402 (0.396-0.408)** | **0.414 (0.408-0.42)** | 0.093 (0.086-0.099) | *0.196 (0.189-0.203)* | *0.154 (0.147-0.161)* | 0.075 (0.067-0.082) | 0.136 (0.129-0.142) | *0.188 (0.181-0.195)* |
| Rank of jfis | | | 1 (1-1) | **0.77 (0.767-0.774)** | **0.894 (0.891-0.896)** | **0.746 (0.742-0.75)** | 0.068 (0.06-0.075) | *0.202 (0.196-0.209)* | *0.175 (0.168-0.182)* | 0.084 (0.076-0.091) | 0.14 (0.132-0.146) | *0.203 (0.196-0.21)* |
| Rank of jfis_0510_12 | | | | 1 (1-1) | **0.685 (0.681-0.69)** | **0.888 (0.885-0.89)** | 0.07 (0.063-0.078) | *0.177 (0.17-0.184)* | *0.159 (0.151-0.165)* | 0.083 (0.076-0.09) | 0.118 (0.111-0.125) | *0.185 (0.178-0.192)* |
| Rank of jcs | | | | | 1 (1-1) | **0.784 (0.781-0.787)** | 0.064 (0.056-0.071) | *0.202 (0.195-0.209)* | *0.185 (0.178-0.191)* | 0.093 (0.085-0.1) | 0.139 (0.132-0.147) | *0.211 (0.205-0.218)* |
| Rank of jcs_0510_12 | | | | | | 1 (1-1) | 0.07 (0.063-0.077) | *0.174 (0.168-0.182)* | *0.159 (0.152-0.167)* | 0.081 (0.073-0.088) | 0.116 (0.108-0.123) | *0.185 (0.178-0.193)* |
| Rank of facebook | | | | | | | 1 (1-1) | 0.074 (0.065-0.082) | 0.091 (0.082-0.099) | 0.098 (0.089-0.107) | 0.101 (0.091-0.11) | **0.283 (0.276-0.29)** |
| Rank of blogs | | | | | | | | 1 (1-1) | 0.097 (0.088-0.106) | *0.158 (0.147-0.169)* | **0.242 (0.231-0.254)** | **0.243 (0.235-0.252)** |
| Rank of twitter | | | | | | | | | 1 (1-1) | 0.132 (0.122-0.141) | 0.139 (0.13-0.148) | *0.906 (0.904-0.908)* |
| Rank of gplus | | | | | | | | | | 1 (1-1) | *0.175 (0.16-0.189)* | *0.197 (0.189-0.205)* |
| Rank of news | | | | | | | | | | | 1 (1-1) | *0.201 (0.193-0.208)* |
| Rank of total altmetrics | | | | | | | | | | | | 1 (1-1) |

**Appendix V. Precision-recall curves for JCS (blue line) and total altmetrics (green line) for identifying 1% most highly cited publications across disciplines: tight analysis**

**General**

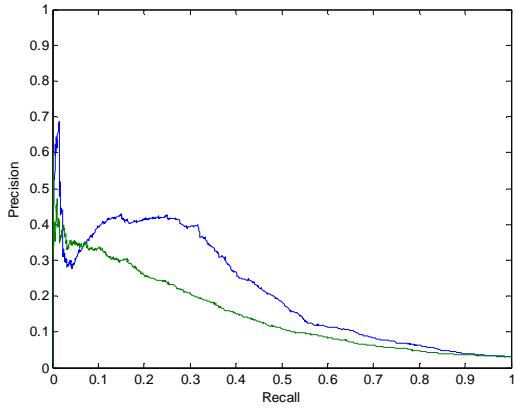

**Biomedical and health sciences**

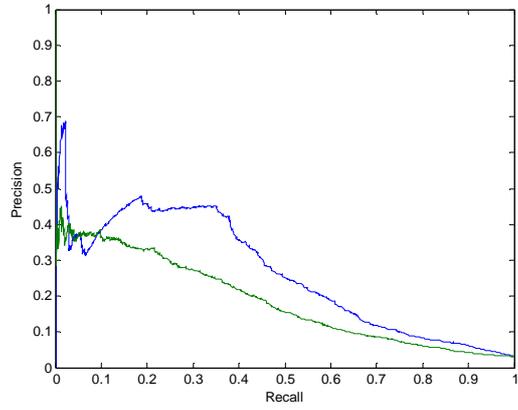

**Life and earth sciences**

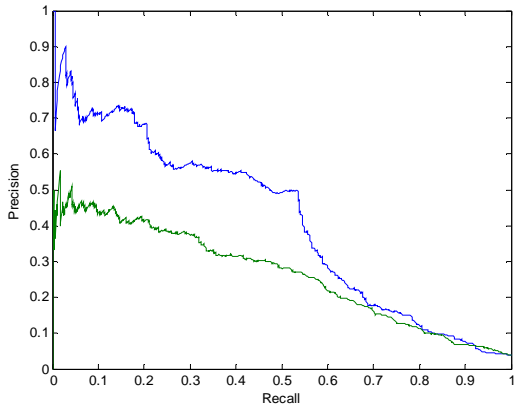

**Mathematics and computer science**

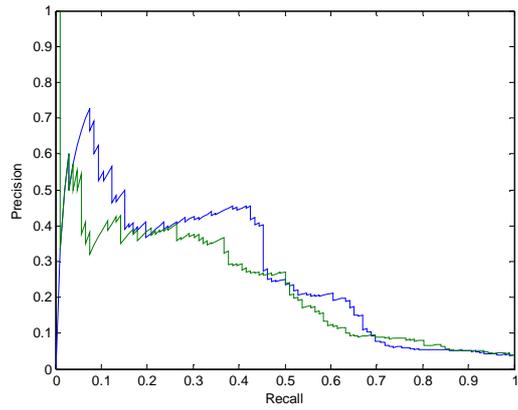

**Natural Sciences and engineering**           **Social sciences and humanities**

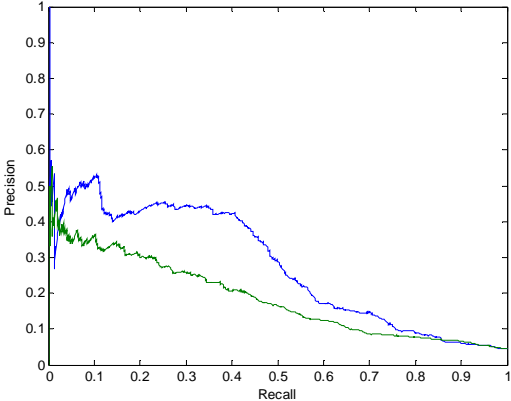
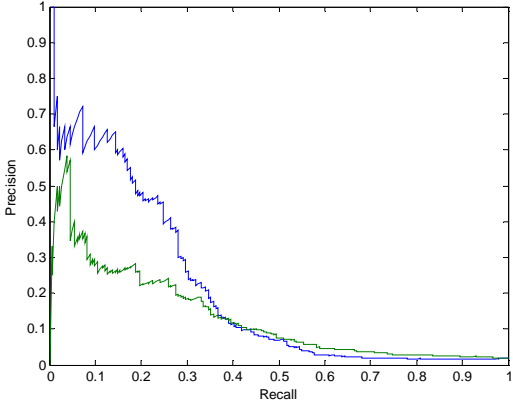